\documentclass[twocolumn,showpacs,preprintnumbers,amsmath,amssymb,prl]{revtex4}

\usepackage{graphicx}
\usepackage{bm}

\begin{document}

\title{Waterlike thermodynamic anomalies in a repulsive-step potential system}

\author{N. V. Gribova}
\affiliation{Institute for High Pressure Physics, Russian Academy
of Sciences, Troitsk 142190, Moscow Region, Russia, and Frankfurt
Institute for Advanced Studies, J.W. Goethe-Universit\"at,
Ruth-Moufang-Str. 1,  D-60438, Frankfurt am Main, Germany}

\author{Yu. D. Fomin}
\affiliation{Institute for High Pressure Physics, Russian Academy
of Sciences, Troitsk 142190, Moscow Region, Russia}

\author{Daan Frenkel}
\affiliation{FOM Institute for Atomic and Molecular Physics,
Amsterdam, The Netherlands and Dept. of Chemistry, Univ. of
Cambridge, Cambridge, UK}

\author{V. N. Ryzhov}
\affiliation{Institute for High Pressure Physics, Russian Academy
of Sciences, Troitsk 142190, Moscow Region, Russia}

\date{\today}

\begin{abstract}
We report a  computer-simulation study of the equilibrium phase
diagram of a three-dimensional system of particles with a
repulsive step potential. The phase diagram is obtained using
free-energy calculations. At low temperatures, we observe a number
of distinct crystal phases. We show that at certain values of the
potential parameters the system exhibits the water-like
thermodynamic anomalies: density anomaly and diffusion anomaly.
The anomalies disappear with increasing the repulsive step width:
their locations move to the region inside the crystalline phase.
\end{abstract}

\pacs{61.20.Gy, 61.20.Ne, 64.60.Kw} \maketitle %

Some liquids (for example, water, silica, silicon, carbon, and
phosphorus) show anomalous behavior in the vicinity of their
freezing lines \cite{deben2003,bul2002,
angel2004,book,book1,deben2001,netz}. The water phase diagrams
have regions where a thermal expansion coefficient is negative
(density anomaly), a self-diffusivity increases upon pressuring
(diffusion anomaly), and the structural order of the system
decreases upon compression (structural anomaly)
\cite{deben2001,netz}. The regions where these anomalies take
place form nested domains in the density-temperature
\cite{deben2001} (or pressure-temperature \cite{netz}) planes:
the density anomaly region is inside the diffusion anomaly
domain, and both of these anomalous regions are inside the
broader structurally anomalous region. In the case of water
these anomalies are usually related to the anisotropy of the
intermolecular potential. However, isotropic potentials are also
able to produce density and diffusion anomalies. It is
interesting that such potentials may be purely repulsive and can
be considered as the simplest models for the water-type
anomalies. It has been shown that water-like structural,
thermodynamic, and dynamic anomalies can be generated in systems
where particles interact via isotropic potentials with two
characteristic length scales, with shorter range corresponding
to a hard-corelike steep repulsion and longer range representing
softer repulsion –- potentials in which two preferable
interparticle distances compete depending on the thermodynamic
conditions of the system
\cite{8,9,10,11,12,13,14,15,16,17,18,19,20,21,22,23,24,25,26}.
In these studies was found that there is an exception case --
the repulsive-step potential -- in which no anomalies were
reported yet \cite{barb2008-1}. In this sense, it is very
interesting to mention the recent study \cite{barb2008} of
evolution of the behavior of the water-like anomalies in the
system with a tunable potential ranging from a ramp potential,
which has all mentioned above anomalies to the repulsive-step
potential, where no anomalies were found so far
\cite{barb2008-1}. In \cite{barb2008} it was shown that
potentials in which two preferred distances are present always
exhibit water-like anomalies, but sometimes they are in an
inaccessible region, as inside a crystal phase. This is the case
for the repulsive-step potential studied in Ref.
[\onlinecite{barb2008-1}].

However, recently it was shown that water-like anomalies can
exist in the systems of particles interacting through the
repulsive step potential \cite{FFGRS2008} for some values of the
potential parameters.

This potential was introduced in the early work of Hemmer and
Stell \cite{8,9} in order to describe isostructural phase
transitions in materials such as Ce or Cs and is the simplest
example of a repulsive intermolecular potential that has a region
of negative curvature in the repulsive part, a feature that is
known to be present in the interatomic potentials of some pure
metallic systems, metallic mixtures, electrolytes and colloidal
systems. Systems of particles interacting through such pair
potentials  can possess a rich variety of phase transitions and
thermodynamic anomalies, including liquid-liquid phase transitions
\cite{RS2002,RS2003,FRT2006}, and isostructural transitions in the
solid region \cite{young1,fren97,stishov}.

In this sense, the purpose of this paper is straightforward. We
will show that the water-like anomalies do exist for the
repulsive step potential, but with increasing the width of the
repulsive step they move to the inaccessible region inside the
crystal phase. The width of the repulsive step of the potential
considered in Refs. [\onlinecite{barb2008-1,barb2008}]
corresponds exactly to this limiting case.

The repulsive step potential has the form:
\begin{equation}
\Phi (r)=\left\{
\begin{array}{lll}
\infty , & r\leq d \\
\varepsilon , & d <r\leq \sigma  \\
0, & r>\sigma%
\end{array}%
\right.  \label{1}
\end{equation}
where $d$ is the diameter of the hard core, $\sigma$ is the width
of the repulsive step,  and  $\varepsilon$ its height. In the
low-temperature limit $\tilde{T}\equiv k_BT/\varepsilon<<1$  the
system reduces to a hard-sphere systems with hard-sphere diameter
$\sigma$, whilst in the limit $\tilde{T}>>1$ the system reduces to
a hard-sphere model with a hard-sphere diameter $d$. For this
reason, melting at high and low temperatures follows simply from
the hard-sphere melting curve $P=cT/\sigma'^3$, where $c \approx
12$ and $\sigma'$ is the relevant hard-sphere diameter ($\sigma$
and $d$, respectively). A changeover from the low-$T$ to high-$T$
melting behavior should occur for $\tilde{T} ={\mathcal O}(1)$.
The precise form of the phase diagram depends on the ratio
$s\equiv \sigma/d$. For large enough values of $s$ one should
expect to observe in the resulting melting curve a maximum that
should disappear as $s\rightarrow 1$ \cite{stishov}. The phase
behavior in the crossover region may be very complex, as  shown in
\cite{FFGRS2008}.

\begin{figure}
\includegraphics[width=7cm]{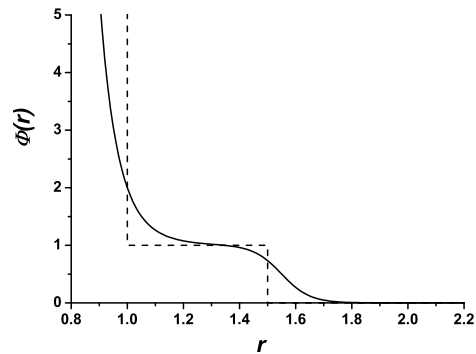}%
\caption{\label{fig:fig1}  A repulsive step potential consisting
of a hard core plus a finite shoulder (dashed line)
($\varepsilon=1, \sigma=1.5$) along with the continuous version of
the potential (\ref{2}) used in the simulations  ($\varepsilon=1,
\sigma_s=1.55$). }
\end{figure}

In our simulations we have used a smoothed version of the
repulsive step potential (Eq.~(\ref{1})), which has the form:
\begin{equation}
\Phi (r)=
\left(\frac{d}{r}\right)^{n}+\frac{1}{2}\varepsilon\left(1-\tanh\left(k_0
\left(r-\sigma_s \right)\right)\right) \label{2}
\end{equation}
where $n=14,k_0=10$. We have considered the following values of
$\sigma_s$: $\sigma_s=1.15, 1.35, 1.55, 1.8$. In
Fig.~\ref{fig:fig1} the repulsive step potential is shown along
with its smooth version which was used in our Monte-Carlo (MC) and
molecular dynamics (MD) simulations.

In the remainder of this paper we use the dimensionless
quantities: $\tilde{{\bf r}}\equiv {\bf r}/d$, $\tilde{P}\equiv P
d^{3}/\varepsilon ,$ $\tilde{V}\equiv V/N d^{3}\equiv
1/\tilde{\rho}$. As we will only use these reduced variables, we
omit the tildes.

In \cite{FFGRS2008} the phase diagrams of the repulsive step
potential system were reported for $\sigma_s=1.15, 1.35, 1.55$. In
the present article we also calculate the phase diagram of the
system for $\sigma_s=1.8$. To determine the phase diagram at
non-zero temperature, we performed constant-NVT MD simulations
combined with free-energy calculations. In all cases, periodic
boundary conditions were used. The number of particles varied
between 250, 500 and 864. No system-size dependence of the results was
observed. The system was equilibrated for $5\times 10^6$ MD time
steps. Data were subsequently collected during $3\times 10^6\delta
t$ where the time step $\delta t=5\times 10^{-5}$.

In order to map out the phase diagram of the system, we computed
its Helmholtz free energy using the thermodynamic integration: the
free energy of the liquid phase was computed via thermodynamic
integration from the dilute gas limit~\cite{book_fs}, and the free
energy of the solid phase was computed by thermodynamic
integration to an Einstein crystal~\cite{book_fs,fladd}.  In the
MC simulations of solid phases, data were collected during $5
\times 10^4$ cycles after equilibration. To improve the statistics
(and to check for internal consistency) the free energy of the
solid was computed at many dozens of different state-points and
fitted to multinomial function. The fitting function we used is
$a_{p,q}T^pV^q$, where $T$ and $V=1/\rho$ are the temperature and
specific volume and powers $p$ and $q$ are connected through
$p+q=N$. The value N we used for the most of calculations is 5.
For the low-density FCC phase  N was taken equal to 4, since we
had less data points. The transition points were determined by a
double-tangent construction.

The region where we have expected thermodynamic anomalies is
situated close to the glassy phase, that means that proper
sampling of the phase space can be problematic. To overcome this
problem we have used the parallel tempering method \cite{book_fs}.
Instead of simulating one system we consider $n$ systems, each
running in the NVT ensemble at a different temperature. Systems at
high temperatures go easily over potential barriers and systems at
low temperatures sample the local free energy minima. The idea of
parallel tempering is to put over MD the MC scheme of
accepting/rejecting a move, but in our case it would be accepting
or rejecting a swap of temperatures between different
configurations after each full (equilibration together with
sampling) MD run. If the low and high temperatures are far apart,
the probability to exchange the configurations is quite low, that
is why we use a range of 'intermediate' temperatures between them
with a small temperature step. So after running the whole parallel
tempering scheme we get a row of systems with subsequent
temperatures and each of the systems was sampled several times.
For our problem we usually used 8 temperatures and tried to swap
them 40 times. This simulation took almost 24 hours running it on
8 processors in parallel at the Joint Supercomputing Center of
Russian Academy of Sciences.

\begin{figure}
\includegraphics[width=7cm]{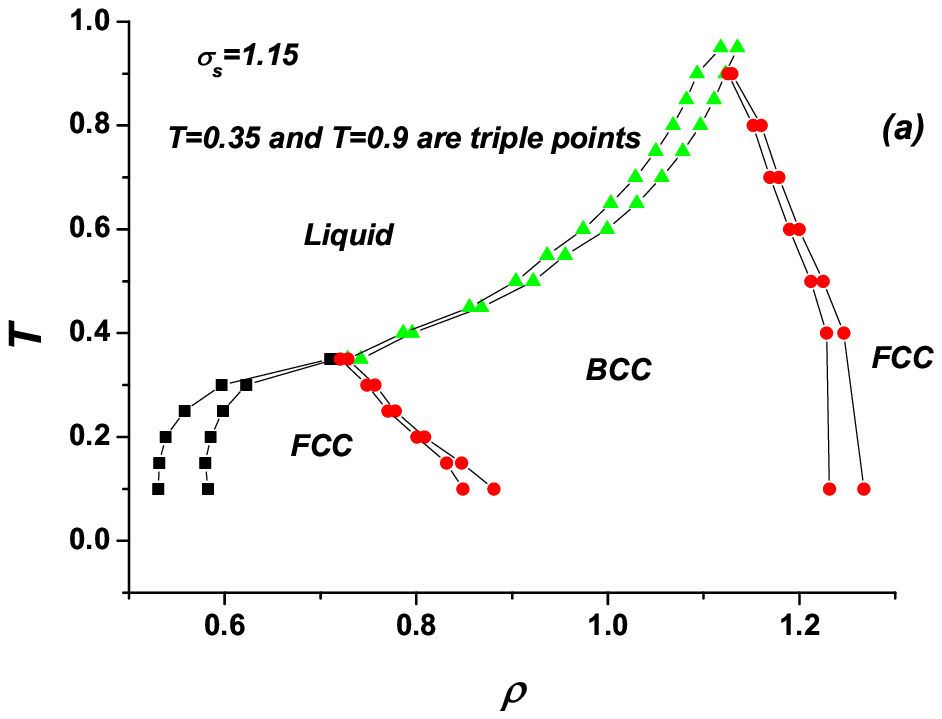}

\includegraphics[width=7cm]{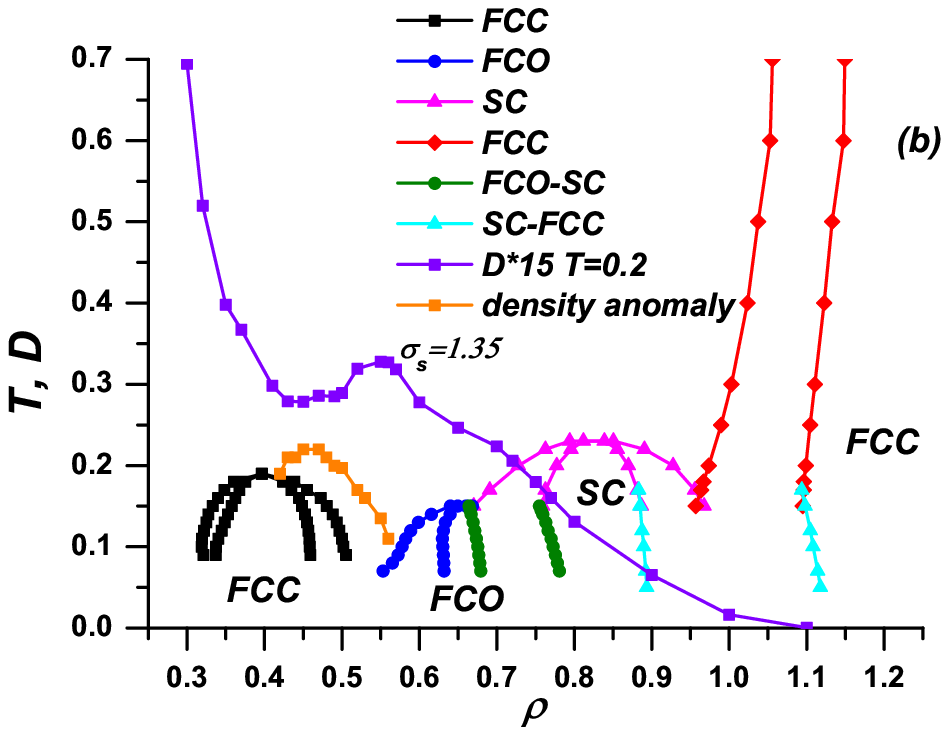}%

\includegraphics[width=7cm]{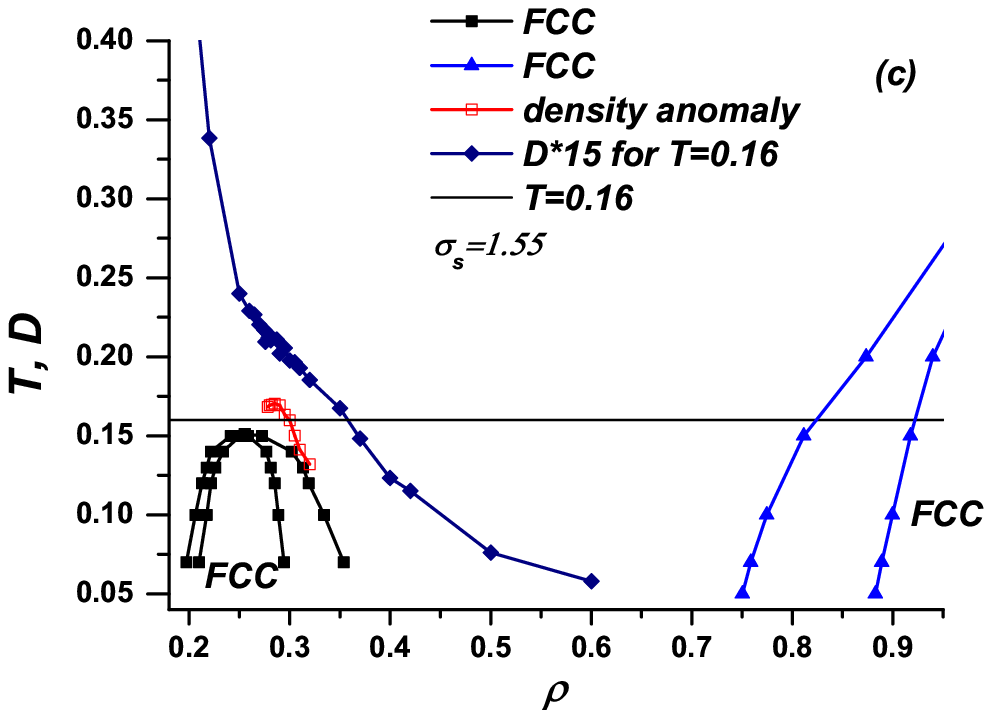}%

\includegraphics[width=7cm]{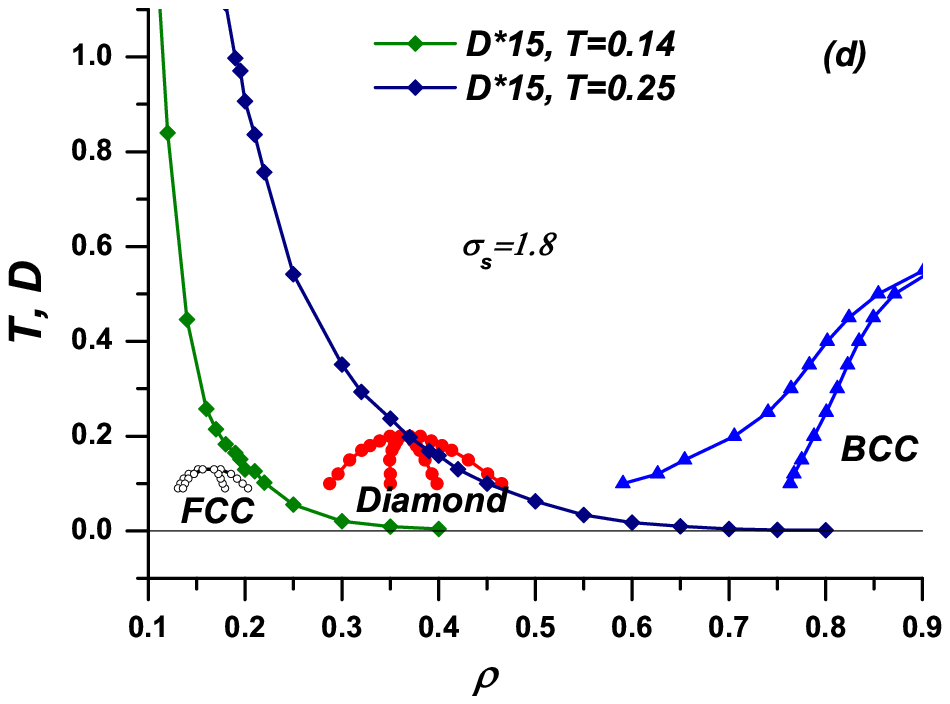}%
\caption{\label{fig:fig2} Phase diagram of the system of
particles interacting through the potential (2) with
$\sigma_s=1.15, 1.35, 1.55, 1.8$ in $\rho-T$ plane. In Figs.2
(b-d) it is shown the behavior of the diffusivity as a function
density. In Figs.2 (b-c) we also represent the locations of the
minima on the isochores (see Fig.~\ref{fig:fig2}.) }
\end{figure}

Fig.~\ref{fig:fig2} shows the phase diagrams that we obtain from
the free-energy calculations for four different values of
$\sigma_s$ (we included the phase diagrams for $\sigma_s=1.15,
1.35, 1.55$ for completeness). Fig.~\ref{fig:fig2}(a) shows the
phase diagram of the system with $\sigma_s=1.15$.  One can see
that for the system with $\sigma_s=1.15$ there are no maxima in
the melting curve. In a soft-sphere system described by the
potential $1/r^{14}$ a face-centered cubic crystal structure has
been reported~\cite{Kofke}. However, the addition of a small
repulsive step leads to the appearance of the FCC-BCC transition
shown in Figs.~\ref{fig:fig2}(a).

Fig.~\ref{fig:fig2}(b) shows the phase diagram of the system
with $\sigma_s=1.35$ in the $\rho-T$ plane. There is a clear
maximum in the melting curve at low densities. The phase diagram
consists in two isostructural FCC parts corresponding to close
packing of the small and large spheres separated by a sequence
of structural phase transitions. This phase diagram was
discussed in detail in our previous publication
\cite{FFGRS2008}. It is important to mention that there is a
region of the phase diagram where we have not found any stable
crystal phase. We think that no crystal structure is stable in
this density range because of frustration as it was discussed in
\cite{FFGRS2008}. In \cite{FFGRS2008} it was shown that the
glass transition occurs in this region with $T_g=0.079$ at
$\rho=0.53$. The apparent glass-transition temperature is above
the melting point of the low-density FCC and FCO phases (see
Fig.~\ref{fig:fig2}(b)). This suggests that  the ``glassy''
phase that we observe is thermodynamically stable. This is
rather unusual for one-component liquids. In simulations, glassy
behavior is usually observed in metastable mixtures, where
crystal nucleation is kinetically suppressed. One could argue
that, in the glassy region, the present system behaves like a
``quasi-binary'' mixture of spheres with diameters $d$ and
$\sigma_s$ and that the freezing-point depression is analogous
to that expected in a binary system with a eutectic point: there
are some values of the diameter ratio such that crystalline
structures are strongly unfavorable and the glassy phase is
stable even for very low temperatures. The glassy behavior in
the reentrant liquid disappears at higher temperatures.

One can expect the frustration to be even more pronounced if we
increase the step size. In Fig.~\ref{fig:fig2}(c) we show the
phase diagram of the system with the potential (\ref{2}) for
$\sigma_s=1.55$. One can see that the system also demonstrates
low- and high density FCC phases separated by FCC to BCC
transitions and the amorphous gap which is much more wider than
for $\sigma_s=1.35$. We did not find any crystal structure
between these isostuctural phases in our study.  The glass
transition temperature is $T_g=0.11091$ at $\rho=0.5$. One can
see that the glassification temperature becomes higher. Given
the lack of crystal structure between crystalline phases and the
increase of the glass transition temperature one can assume that
the frustration effects become higher with the increase of the
step width.

\begin{figure}
\includegraphics[width=7cm]{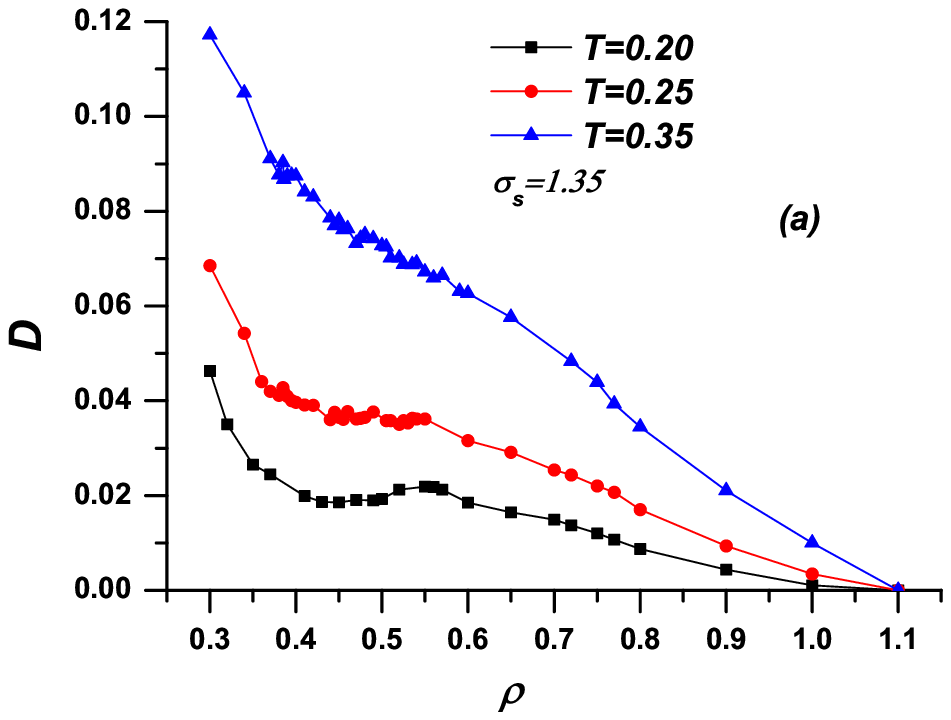}%

\includegraphics[width=7cm]{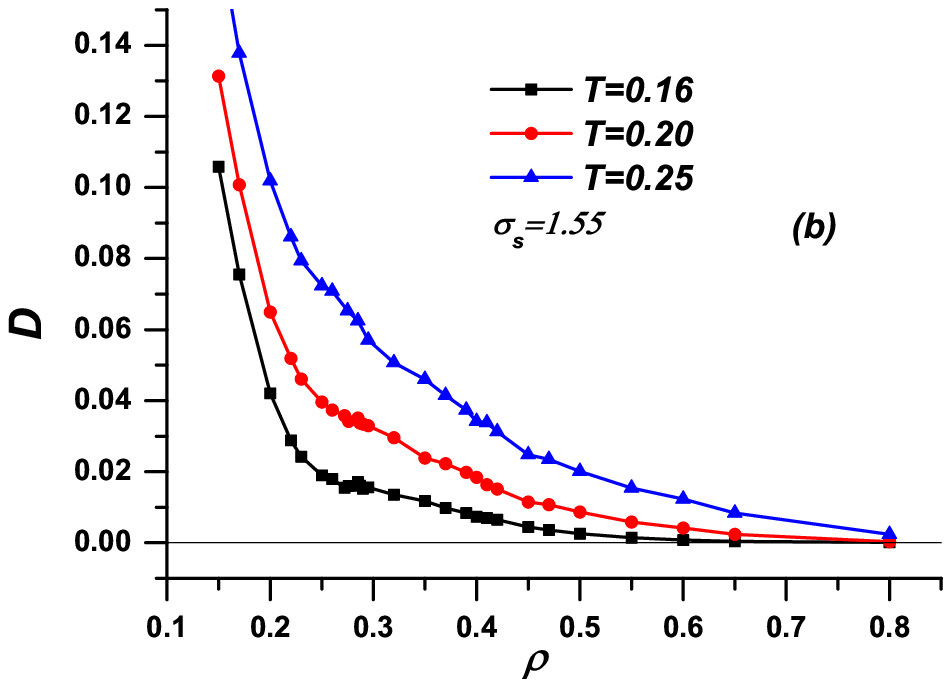}%

\includegraphics[width=7cm]{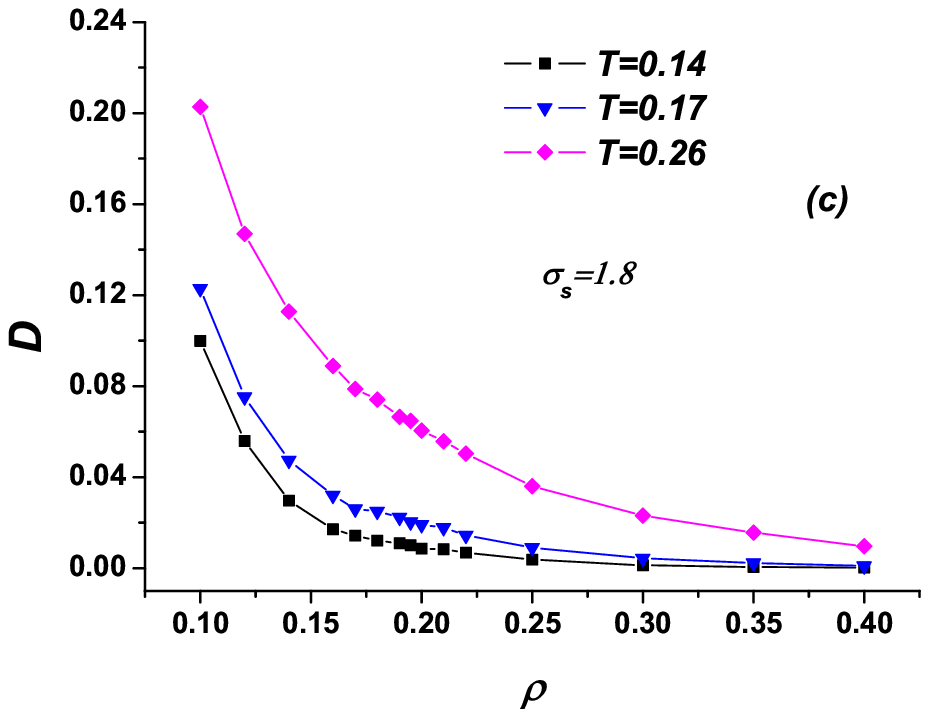}%
\caption{\label{fig:fig3} Diffusion anomaly for $\sigma=1.35,
1.55, 1.8$.  }
\end{figure}

The phase diagram of the system with $\sigma_s=1.8$ is shown in
Fig.~\ref{fig:fig2}(d). One can see that inside the disordered
gap in the phase diagram there appears the crystalline phase
with diamond structure, however, this phase does not extend over
the whole disordered region in the phase diagram.

\begin{figure}
\includegraphics[width=7cm]{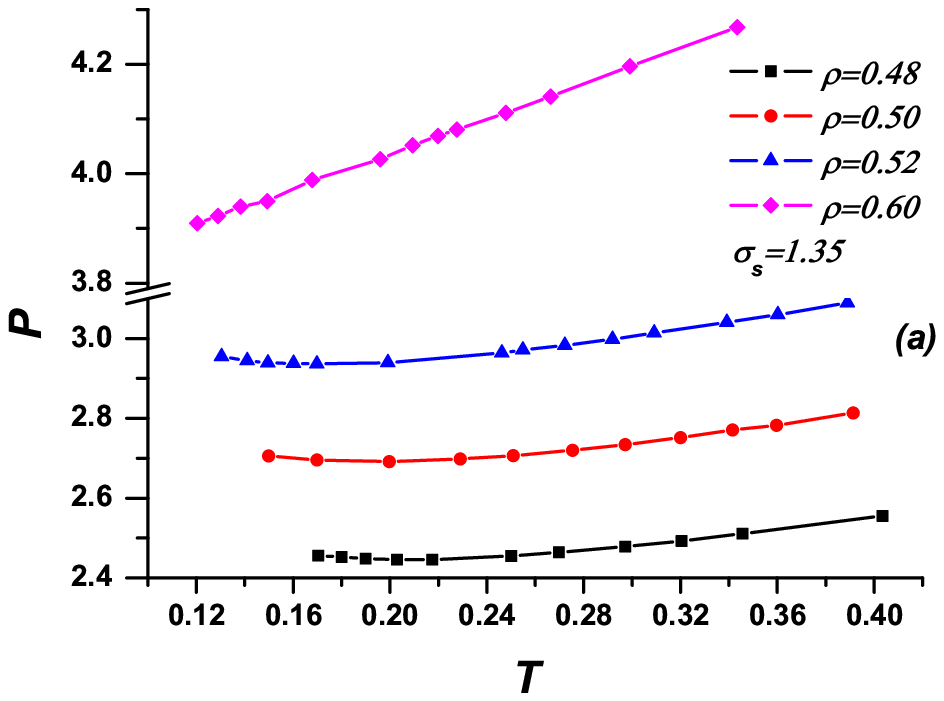}%

\includegraphics[width=7cm]{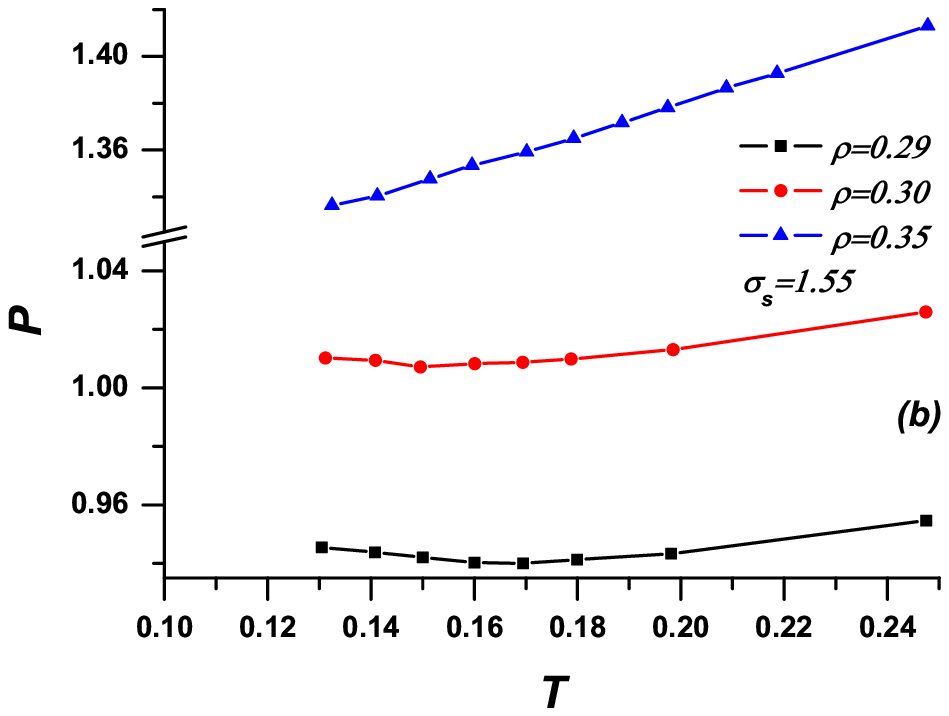}%

\includegraphics[width=7cm]{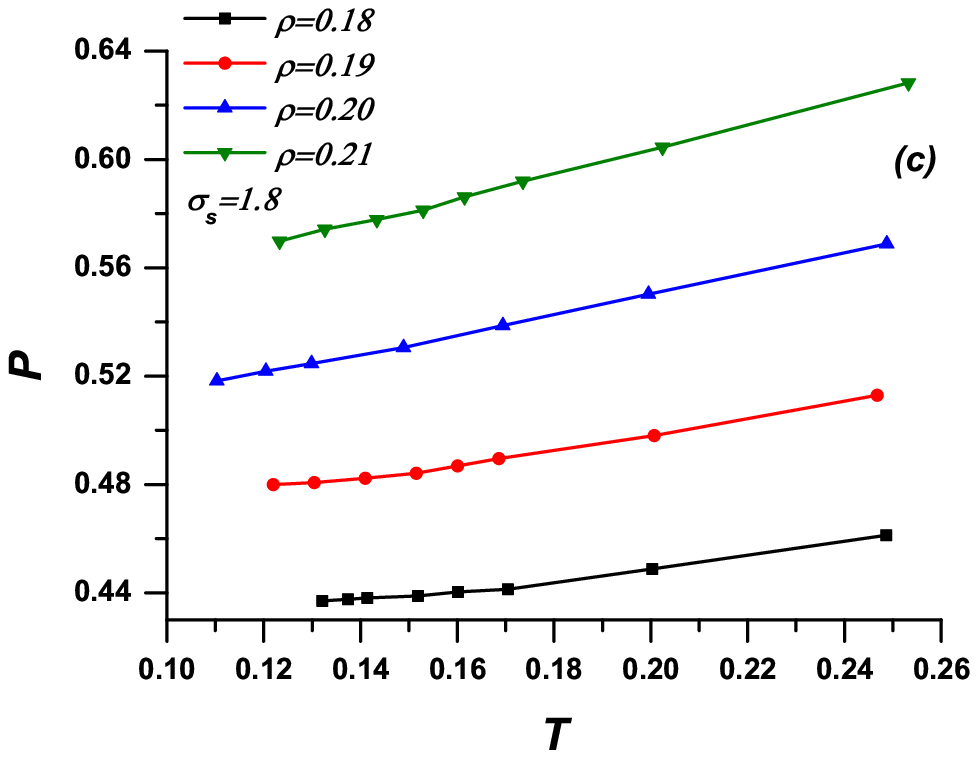}%
\caption{\label{fig:fig4} Density anomaly for $\sigma=1.35, 1.55,
1.8$. }
\end{figure}

As it was mentioned above, one can expect the appearance of
thermodynamic anomalies in the vicinity of the anomalous points
on the phase diagrams of the repulsive-step potential system. To
check this point, we calculated the isochores and diffusivity
for different values of $\sigma_s$. In this sense, it is not
surprising that there are no thermodynamic anomalies for
$\sigma_s=1.15$. It is known, that for normal liquids the
diffusivity decreases monotonically with increasing density at
constant temperature. In contrast, we have observed in our
model, that for a certain values of the potential parameters,
for the densities in the vicinity and above the maximum of the
melting curve, the diffusivity curve has a bend (see
Figs.~\ref{fig:fig3}(b-c)). In Figs.~\ref{fig:fig3} the behavior
of the diffusivity is shown in more detail for different values
of $\sigma_s$. One can see that with increasing the width of the
repulsive step $\sigma_s$ the anomaly is becoming less
pronounced and disappears for $\sigma_s=1.8$. It is interesting
to note that this value of $\sigma_s$ corresponds to width of
the repulsive step considered in  \cite{barb2008-1,barb2008}
where  no anomalies were found for the repulsive step potential.

The region where the diffusivity anomaly exists almost coincides
with a region in which the isochore has a minimum instead of
growing monotonically (see Figs.~\ref{fig:fig2}(b-c), where the
locations of the minima of isochores are shown, and
Figs.~\ref{fig:fig4}). Using the thermodynamic relation
$\left(\partial P/\partial T\right)_V=\alpha_P/K_T$, where
$\alpha_P$ is a thermal expansion coefficient and $K_T$ is the
isothermal compressibility and taking into account that $K_T$ is
always positive and finite for systems in equilibrium not at a
critical point, one can conclude that there is a range of
densities and temperatures where the thermal expansion
coefficient $\alpha_P$ is negative.

\begin{figure}
\includegraphics[width=7cm]{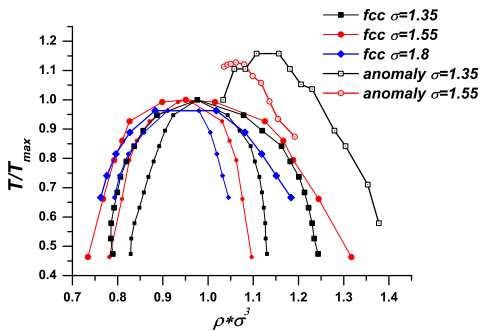}%
\caption{\label{fig:fig5} Re-scaled part of the phase diagram
with locations of the minima on the isochores.}
\end{figure}

To elucidate the behavior of the anomalies with increasing the
width of the repulsive step of the potential, we re-scaled the
parts of the phase diagrams corresponding to the first maximum
on the melting curve (see Fig.~\ref{fig:fig2}) by multiplying
the density by the $\sigma_s^3$ and dividing the temperature by
$T_{max}$, where $T_{max}$ is the temperature corresponding to
the maximum. In accordance with the qualitative picture depicted
above the re-scaled parts of the phase diagrams should coincide.
As it is seen in Fig.~\ref{fig:fig5} this is approximately the
case for $\sigma_s=1.35, 1.55, 1.8$. The discrepancies between
the curves appear to be because we consider the smoothed version
of the repulsive step potential. In Fig.~\ref{fig:fig5} we also
show the locations of the minima of the isochores for
$\sigma_s=1.35, 1.55$. One can see that with increasing the
width of the repulsive step the line of the minima moves to the
melting line and becomes invisible in the metastable region. It
should be noticed that this scenario is similar to the one
depicted in Ref.~\onlinecite{barb2008}.

At low densities, we have effectively a liquid consisting of
spheres with diameter $\sigma_s$, at high densities, the liquid
consists of spheres with diameter $d$. In the ``anomalous region''
inbetween, our system appears as a mixture of both sorts of
particles, and one can expect that in this region structural order
should decrease for intermediate values of $\sigma_s$. In this
case, the entropy of the system should increase with increasing
density, and, due to the thermodynamic relation
$\left(\frac{\partial\rho}{\partial T}\right)_P=\rho^2
\left(\frac{\partial\rho}{\partial P}\right)_T
\left(\frac{\partial s}{\partial \rho}\right)_T$ \cite{deb2005},
one gets the anomalous behavior in this region. This further
demonstrates that our model shows a quasi-binary behavior.


In summary, we have performed the extensive computer simulations
of the phase behavior of  systems described by the soft, purely
repulsive step potential (\ref{2}) in three dimensions.  We find
a surprisingly complex phase behavior. We argue that the
evolution of the phase diagram  may be qualitatively understood
by considering this one-component system as a quasi-binary
mixture of large and small spheres. Interestingly,  the phase
diagram includes two crystalline FCC domains separated by a
sequence of the structural phase transitions and a reentrant
liquid that becomes amorphous at low temperatures. The
water-like anomalies (density anomaly and diffusion anomaly)
were found in the reentrant liquid for $\sigma_s=1.35, 1.55$.
The anomalies disappear with increasing the  repulsive step
width: their locations move to the region inside the crystalline
phase in the vicinity of the maximum on the melting line.

\begin{acknowledgments}
We thank S. M. Stishov and V. V. Brazhkin for stimulating
discussions.
N.G. thanks A. Arnold for introduction to parallel computing and
valuable remarks. N.G. and Y.F. thank the Joint Supercomputing
Center of Russian Academy of Sciences for computational power.
The work was supported in part by the Russian
Foundation for Basic Research (Grant No 08-02-00781) and the
Fund of the President of Russian Federation for Support of Young
Scientists (MK-2905.2007.2).

\end{acknowledgments}




























\end{document}